# The new concept of nano-device spectroscopy based on Rabi-Bloch oscillations for THz-frequency range

**Ilay Levie, and Gregory Slepyan**

School of Electrical Engineering, Tel Aviv University, Tel Aviv, Israel

**Abstract:** We considered one-dimensional quantum chains of two-level Fermi particles coupled via the tunneling driven both by ac and dc fields in the regimes of strong and ultrastrong coupling. The frequency of ac field is matched with the frequency of the quantum transition. Based on the fundamental principles of electrodynamics and quantum theory, we developed a general model of quantum dynamics for such interactions. We showed that the joint action of ac and dc fields leads to the strong mutual influence of Rabi- and Bloch oscillations one to another. We focused on the regime of ultrastrong coupling, for which Bloch- and Rabi-frequencies are a significant values of the frequency of interband transition. The Hamiltonian was solved numerically with account of anti-resonant terms. It manifests by the appearance of great number of narrow high-amplitude resonant lines in the spectra of tunneling current and dipole moment. We proposed the new concept of THz spectroscopy promising for different applications in future nanoelectronics and nano-photonics.

1. Introduction

Since the early days of physics, spectroscopy has been a platform for many fundamental investigations. One of the central concepts in spectroscopy is a resonance with its corresponding resonant frequency. Types of spectroscopy can also be distinguished by the nature of the interaction between the electromagnetic field (EM-field) and the condensed matter (as examples, one may mention absorption and emission of light, Raman and Compton scattering, nuclear magnetic resonance, etc.). The first period in the development of spectroscopy is associated with atomic and molecular spectroscopy. Atoms of different elements have distinct spectra, and therefore atomic spectroscopy allows for the identification of a sample's elemental composition. The spectral lines obtained by experiment and theory can be mapped to the Mendeleev Periodic Table of the Elements. Spectroscopic studies at this time were central to the development of quantum mechanics.

The second period in the making of spectroscopy is associated with a wide range of applications to different types of chemical substances: gases, liquids, crystals, polymers, etc. The combination of atoms or molecules into macroscopic samples leads to the creation of many-particle quantum states, whereby different materials have distinct spectra. Therefore, spectroscopy observations in a wide frequency range (from microwave until gamma-ray) became irreplaceable tool for the identification of a sample's structure, chemical composition, samples quality, etc. in material engineering, physical chemistry and biophysics. Recent progress in electronics and informatics is associated with the development of nanotechnologies and synthesis of different types of nano-objects. The distinct spectra became the attribute of nano-objects of various spatial configurations, but the same chemical composition. In this context, the future period in the spectroscopy development would be probably associated with the spectroscopy of the whole range of electronic devices or their rather large components. This trend will provide tools of the tunable spectroscopy adapted for such types of tasks. The promising examples from this point of view are based on the strong and ultrastrong interactions of condensed matter with EM-field.

The strong-coupling field-matter interaction is defined as a regime, in which the EM-field does not correspond the small perturbation [1]. It leads to the periodical transitions of a two-level quantum system between its stationary states under the action of ac driving field (Rabi-oscillations – RO) [1]. The phenomenon was theoretically predicted by Rabi on nuclear spins in radio-frequency magnetic field [2] and afterwards, discovered in various physical systems, such as electromagnetically driven atoms [3], semiconductor quantum dots [4] and different types of solid-state qubits [5-9]. The simplest physical picture of Rabi effect is given by the model of single atom [1]. It can be essentially modified by a set of additional features, such as the broken inversion symmetry [10], the spatial RO propagation in the arrays of coupled Rabi oscillators (Rabi-waves) and local field depolarization effect [11-16].

The early quantum theory of electrical conductivity in crystal lattices by Bloch, Zener and Wannier [17-20] led to the prediction that a homogeneous dc field induces an oscillatory rather than uniform motion of the electrons. These so-called Bloch oscillations (BO) have never been observed in natural crystals because the scattering time of the electrons by the lattice defects is much shorter than the Bloch period [21]. In semiconductor superlattices the larger spatial period leads to a much shorter Bloch period, and BO have been observed through the THz radiation of the electrons [22]. BO in dc biased lattices is due to wave interference and have been observed in a number of quite different physical systems: a few interacting atoms in optical lattices [23,24], ultracold atoms [25–30], light intensity oscillations in waveguide arrays [31-37], acoustic waves in layered and elastic structures [38], atomic oscillations in Bose-Einstein condensates [39], among others. Several recent studies have investigated the dynamics of cold atoms in optical lattices subject to ac forcing; the theoretically predicted renormalization of the tunneling amplitudes has been verified experimentally. Recent observations include global motion of the atom cloud, such as giant "super–Bloch oscillations" [40]. As a result, BO was transformed from partial physical effect to the universal phenomenon of oscillatory motion of wave packets placed in a periodic potential when driven by a constant force [24, 41].

Recently, there have been several theoretical studies of the ultrastrong light-matter interaction regime [42-45] where the Rabi frequency becomes comparable with the frequency of the inter-zone quantum transition [43]. This regime has been realized experimentally with intersubband transitions coupled with plasmon waveguides in the infrared frequency range [46] and metallic microcavities in the THz [47–49], magnetoplasmons of two-dimensional electron gas [50], superconducting qubits [51], and molecular transitions [52]. It was shown, that the system behaves like a multi-level quantum bit with non-monochromatic energy spacing, owing to the inter-particle interactions.

One of the general tendencies in modern physics consists in the synthesis of different physical mechanisms in the network of one physical process with their strong mutual influence. One of such examples have been demonstrated in [53], where considered the quantum chain of the coupled two-level Fermion systems driven simultaneously by ac and dc fields. It was shown, that in the case of resonant interaction with ac-component the particle dynamics exhibits itself in the oscillatory regime, which may be interpreted as a combination of RO and BO with their strong mutual influence. This type of quantum dynamics was named in [53] Rabi-Bloch oscillations (RBO). Such scenario dramatically differs from the individual picture both types of oscillations due to the interactions. This novel effect is counterintuitive because of the strongly different frequency ranges for such two types of oscillations existence.

In this paper we consider the RBO in the regimes of strong and ultrastrong coupling. We present approximate analytic solution and results of the numerical modelling. We study both the temporal dynamics and spectra, identify the spectral lines, and compare the spectra behavior in the strong and ultrastrong regimes. Conclusively, we discuss the promising applications of our results in the novel types of spectroscopy.

## 2. Structures and models

The model system for RBO observing in the strong coupling regime is shown on Figure 1. It corresponds to the periodic two-level atomic chain, excited with obliquely incident ac field in the strong coupling regime and driven with dc voltage applied at the ends. The neighboring atoms are coupled via interatomic tunneling with different values of penetration at the ground and excited states.

The model system for RBO analysis in the ultrastrong coupling regime is shown on Figure 2. Analysis of RO in the regime of ultrastrong coupling was given in [43] performing on a model system depicted in Figure 2a, where a mode of microcavity of THz frequency range is coupled with the electronic excitation of a highly doped quantum well (QW) inserted inside the capacitor. There are at least two confined subbands in the QW, and the fundamental subband hosts a two-dimensional electron gas with an arbitrary total number of electrons. The energy difference between the first and the second subband is matched with the resonance of the LC circuit. The reduction of the capacitor armatures also implies a reduction of the capacitance C, which in turn changes the resonant frequency of the LC circuit. In order to keep the resonator frequency always matched with the frequency of the quantum transition, one has to increase the inductance L. The strong reduction of the capacitor volume in the quantum LC resonator allows for the realization of the ultrastrong coupling regime with a few electrons only, even with a one electron. It was focused in [43] on the case of electron gas coupled with the circuit photons of the capacitor. One of the peculiar features of such quantum mechanical system is the existence of entangled light-matter excitations in analogy with the super-radiant and sub-radiant Dicke-states in atomic clouds [43].

The system under consideration in our case is in essence different. We consider the periodic chain of elements [43], every one of which plays role of the single atom in Figure1. The chain is driven by both dc and classical THz frequency field (Figure 2b). The THz oscillating voltage is applied to the system through a nano-antenna [54-63] (Figure 2b). The QW assumed to be periodically doped, thus the additional periodic potential has been induced (Figure 2c). We assume only one electron in the system. The coupling is governed by the inter-barrier electron tunneling. In our case, QW represents the effective super-lattice, which is able to support the electron's BO [21] and was the first system in which BO was experimentally observed [21,41].

Of course, these two models are identical from physical point of view. The difference is only in the reachable value of the coupling factor, which makes significant the contribution of anti-resonant terms. These models relate to the different frequency ranges (optical – Figure 1, THz – Figure 2). As it leads from analysis [43], the ultrastrong coupling becomes implementable in THz, while remains open in optics. Thus, the optical model will be simplified via RWA, while the THz one is based on the total Hamiltonian.

We use for analysis the method of probability amplitudes [1]. The single-electron motion is described by the wave function $|\Psi(t)\rangle$ satisfied the Schrodinger equation, $i\hbar \partial_t |\psi\rangle = \hat{H}|\psi\rangle$ where $\hat{H}$ is the total Hamiltonian of the system. The Hamiltonian may be decomposed as $\hat{H} = \hat{H}_0 + \hat{H}_I$, where $\hat{H}_0$ is the component of free electron movement in the chain associated with the tunneling and described by the periodic superlattice potential $V(\mathbf{r})$ (Figure 2c). The second component $\hat{H}_I$ corresponds to the interaction of electron with total (dc and ac) EM field. The electron interaction with EM-field is described in the dipole approximation [1], thus $\hat{H}_I = -e\mathbf{E}(\mathbf{r},t) \cdot \hat{\mathbf{r}}$, $\hat{\mathbf{r}}$ is the operator of electron position in the chain, $\mathbf{E}(\mathbf{r},t) = \mathbf{E}_{Ac}(\mathbf{r},t) + \mathbf{E}_{Dc}(\mathbf{r})$ is the superposition of dc and ac components. Our analysis will base on the Wannier basis [17-20]; the presentation of Hamiltonian in the Wannier basis is given in [53].

.

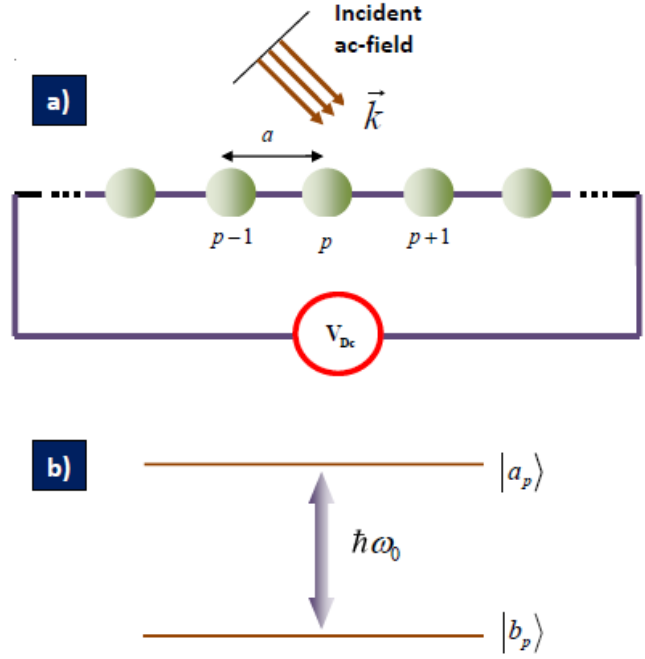

**Figure 1.** a) General illustration of the periodic two-level atomic chain used as a model, indicating RBO in strong coupling regime. It is excited with obliquely incident ac field and driven with dc voltage applied at the ends. The neighboring atoms are coupled via interatomic tunneling with different values of penetration at the ground and excited states. b) Ground and excited energy levels of single two-level atom separated by the transition energy $\hbar\omega_0$.

## 3. Strong coupling regime: analytical approximation

The wave function for the single-particle state of Hamiltonian is given by

$$|\Psi(t)\rangle = \sum_p \left(a_p(t)|a_p\rangle + b_p(t)|b_p\rangle\right) \quad (1)$$

where $|a_p\rangle, |b_p\rangle$ are Wannier functions of the atom with number $p$ in the excited and ground states, respectively [53], $a_p(t), b_p(t)$ the unknown variables, which are satisfy the system differential equations

$$i\frac{\partial a_p}{\partial t} = \left(\frac{\omega_0}{2} - p\Omega_B\right)a_p + t_a\left(a_{p+1} + a_{p-1}\right) - \Omega_R \cos\omega t \cdot b_p \quad (2)$$

$$i\frac{\partial b_p}{\partial t} = -\left(\frac{\omega_0}{2} + p\Omega_B\right)b_p + t_b\left(b_{p+1} + b_{p-1}\right) - \Omega_R \cos\omega t \cdot a_p \quad (3)$$

In this section we will consider equations (2),(3) analytically assuming for simplicity exact resonance condition to be fulfilled ($\omega = \omega_0$).

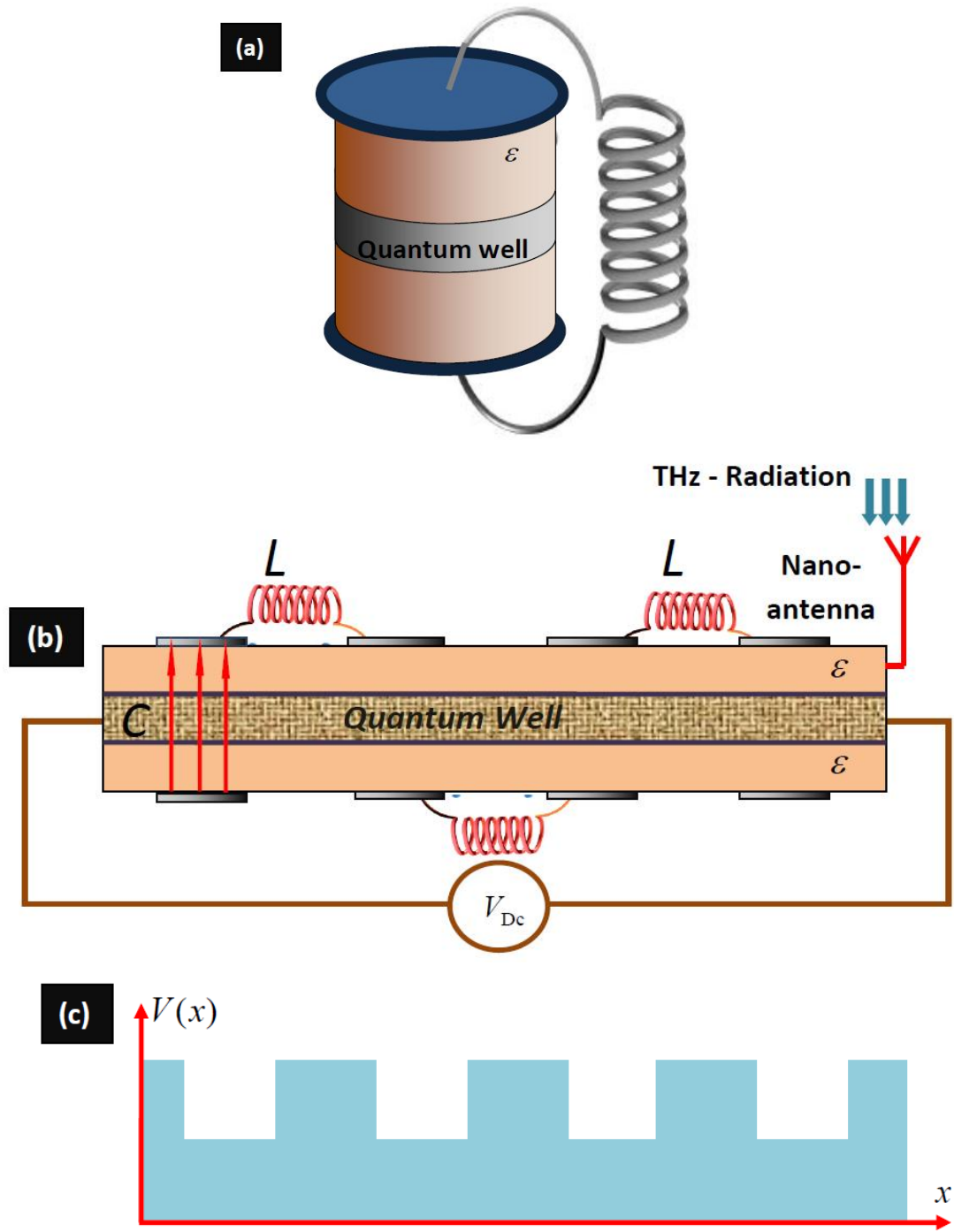

**Figure 2.** General illustration of the periodic chain used as a model, indicating RBO in ultrastrong coupling regime. a) The single element of the chain. It corresponds to a highly doped QW inserted inside the capacitor of microcavity of THz frequency range [43]. It is resonantly coupled with a mode of microcavity. The frequency matching is reached via the choice of inductance value. b) A highly doped QW inserted inside the capacitors of the chain of coupled microcavities. The dc voltage applied to the system at the ends of the chain. The THz oscillating voltage is applied through a nano-antenna. c) The periodic potential induced by additional doping for transforming the homogeneous QW to the effective superlattice. The coupling is governed by the inter-barrier electron tunneling.

## 3.1. Travelling Rabi-waves

Assuming the driven dc field to be absent ($\Omega_B = 0$), we have from (2),(3)

$$i\frac{\partial a_p}{\partial t} = \frac{\omega}{2}a_p + t_a\left(a_{p+1} + a_{p-1}\right) - \Omega_R \cos \omega t \cdot b_p \tag{4}$$

$$i\frac{\partial b_p}{\partial t} = -\frac{\omega}{2}b_p + t_b\left(b_{p+1} + b_{p-1}\right) - \Omega_R \cos \omega t \cdot a_p \tag{5}$$

The idea of simplifying the equations (4),(5) is based on the, so called, rotation-wave approximation (RWA), which takes into account only the co-rotating field with the system and neglects the counter-rotating part [1]. It corresponds to the exchanges $b_p \cos \omega t \to (1/2) b_p \exp(-i\omega t)$, $a_p \cos \omega t \to (1/2) a_p \exp(i\omega t)$ and leads to the equations

$$i\frac{\partial a_p}{\partial t} = \frac{\omega}{2}a_p + t_a\left(a_{p+1} + a_{p-1}\right) - \frac{1}{2}\Omega_R e^{-i\omega t} b_p \tag{6}$$

$$i\frac{\partial b_p}{\partial t} = -\frac{\omega}{2}a_p + t_b\left(b_{p+1} + b_{p-1}\right) - \frac{1}{2}\Omega_R e^{i\omega t} a_p \tag{7}$$

The system (6),(7) is a generalization of Jaynes-Cummings model for the single atom widely used in quantum optics [1], to the case of atomic chain.

The ansatz for solution of equations (6), (7) in the form of travelling wave is

$$\begin{pmatrix} a_p(t) \\ b_p(t) \end{pmatrix} = e^{i(p\varphi - \nu t)} \begin{pmatrix} \tilde{a} e^{-i\frac{1}{2}\omega t} \\ \tilde{b} e^{i\frac{1}{2}\omega t} \end{pmatrix} \tag{8}$$

where $\tilde{a}, \tilde{b}$ are unknown coefficients, $\varphi$ is the phase shift per unit cell, $\nu$ is the eigen-frequency. Substituting (8) to (6),(7), we obtain the system with respect to $\tilde{a}, \tilde{b}$:

$$\begin{pmatrix} \nu - 2t_a \cos\varphi & \Omega_R/2 \\ \Omega_R/2 & \nu - 2t_b \cos\varphi \end{pmatrix} \cdot \begin{pmatrix} \tilde{a} \\ \tilde{b} \end{pmatrix} = 0 \tag{9}$$

Setting determinant equal to zero, we obtain the characteristic equation with respect to $\nu$:

$$(\nu - 2t_a \cos\varphi)(\nu - 2t_b \cos\varphi) - \frac{\Omega_R^2}{4} = 0 \tag{10}$$

with two solutions

$$v_{1,2}(\varphi) = \left\{ (t_a + t_b)\cos\varphi \pm \sqrt{(t_a - t_b)^2 \cos^2\varphi + \frac{\Omega_R^2}{4}} \right\} \tag{11}$$

The linearly independent eigen modes correspondent to eigen frequencies (11) are

$$|\Psi_1(t)\rangle = \frac{1}{\sqrt{N(1+\Lambda^2)}} \sum_p e^{i(p\varphi - v_1 t)} \left( |a_p\rangle e^{-i\frac{1}{2}\omega t} - \Lambda |b_p\rangle e^{i\frac{1}{2}\omega t} \right) \tag{12}$$

$$|\Psi_2(t)\rangle = \frac{1}{\sqrt{N(1+\Lambda^2)}} \sum_p e^{i(p\varphi - v_2 t)} \left( \Lambda |a_p\rangle e^{-i\frac{1}{2}\omega t} + |b_p\rangle e^{i\frac{1}{2}\omega t} \right) \tag{13}$$

where

$$\Lambda = \frac{\Omega_R}{2\left[(t_a - t_b)\cos\varphi + \sqrt{(t_a - t_b)^2 \cos^2\varphi + \Omega_R^2/4}\right]} \tag{14}$$

The Rabi-oscillations propagated along the chain in the form of travelling waves are given by

$$|\Psi(t)\rangle = \frac{1}{\sqrt{N(1+\Lambda^2)}} \sum_p \left( A_p(t) |a_p\rangle + B_p(t) |b_p\rangle \right) e^{-i\mu t} \tag{15}$$

where

$$A_p(t) = \left( C_1 e^{-i\frac{1}{2}\Omega t} + \Lambda C_2 e^{i\frac{1}{2}\Omega t} \right) e^{-i\frac{1}{2}\omega t} e^{ip\varphi} \tag{16}$$

$$B_p(t) = \left( C_2 e^{i\frac{1}{2}\Omega t} - \Lambda C_1 e^{-i\frac{1}{2}\Omega t} \right) e^{i\frac{1}{2}\omega t} e^{ip\varphi} \tag{17}$$

$C_{1,2}$ are arbitrary integration constants satisfied the normalization condition $|C_1|^2 + |C_2|^2 = (1+\Lambda^2)^{-1}$ (guarantied the correct normalization for the wavefunction (15)), $\mu = (t_a + t_b)\cos\varphi$. Let us assume the system initially prepared in the excited state ($B_p(0) = 0$). In this case we obtain from (16),(17)

$$A_p(t) = \frac{1}{1+\Lambda^2} \left( e^{-i\frac{1}{2}\Omega t} + \Lambda^2 e^{i\frac{1}{2}\Omega t} \right) e^{-i\frac{1}{2}\omega t} e^{ip\varphi} \tag{18}$$

$$B_p(t) = \frac{2i\Lambda}{1+\Lambda^2} \sin\frac{\Omega t}{2} e^{i\frac{1}{2}\omega t} e^{ip\varphi} \tag{19}$$

where the RO frequency is given by

$$\Omega(\varphi) = \sqrt{4(t_a - t_b)^2 \cos^2\varphi + \Omega_R^2}. \tag{20}$$

It is important to note that it is corrected with respect to Rabi-frequency due the interatomic coupling via the tunneling.

## 3.2. Tunneling current

Following the general presentation of tunneling current in terms of probability amplitudes [1], we have for RO

$$J_{T,p} = i\frac{e}{2}\left[t_a\left(A_{p+1} - A_{p-1}\right)A_p^* + t_b\left(B_{p+1} - B_{p-1}\right)B_p^*\right] + c.c. =$$
$$= -2e\sin\varphi\left(t_a|A_p|^2 + t_b|B_p|^2\right) \quad (21)$$

For the case of completely excited initial state, we obtain using (18),(19)

$$J_{T,p} = -\frac{2e\sin\varphi}{\left(1+\Lambda^2\right)^2}\left[t_a\left(1+\Lambda^4 + 2\Lambda^2\cos\Omega t\right) + 4t_b\Lambda^2\sin^2\frac{\Omega t}{2}\right] \quad (22)$$

In this case the tunnel current consists of two components: dc-current and the component on the RO frequency (20). Its physical origin is similar to the so-called low-frequency RO in the two-level quantum systems with broken inversion symmetry [10]. As it was predicted in [10], it leads to non-zero diagonal matrix elements of dipole moment. It is impossible for real atoms (excluding hydrogen), but implementable for artificial atoms (for example, self-organized semiconductor quantum dots). As a result, the inter-level optical transitions are accompanied by the intra-level periodical movement with Rabi-frequency, which is responsible for low-frequency radiation. It exists only in the case of asymmetry, which manifests itself in the non-identity of diagonal dipole matrix elements. In our case, the intra-band movement and correspondent asymmetry are associated with interatomic tunneling and different penetration values ($t_a \neq t_b$), respectively.

## 3.3. Rabi-Bloch oscillations: quasi-classical model

BO of the charge particles is a pure quantum effect which can be explained in a simple quasiclassical model [21,24,25]. The similar concept may be developed for quasi-particles (Rabi-oscillated electrons in our case). The periodicity of the lattice leads to a band structure of the energy spectrum and the corresponding eigenenergies $\hbar v_{1,2}(\varphi)$ given by equation (11). The eigenstates $|\Psi_{1,2}(h)\rangle$ given by equations (12), (13) are labeled by the the continuous quasimomentum $h$; $\hbar v_{1,2}(h,n)$ and $|\Psi_{1,2}(h)\rangle$ are periodic functions of $h$ with period $2\pi/a$, and $h$ is conventionally restricted to the first Brillouin zone $[-\pi/a; \pi/a]$. Under the influence of dc field, weak enough not to induce interband transitions, a given state $|\Psi_{1,2}(h_0)\rangle$ evolves up to a phase factor into the state $|\Psi_{1,2}(h(t))\rangle$ with $h(t)$ variation according to

$$\frac{dh}{dt} = -\frac{eE_{Dc}}{\hbar} \quad (23)$$

or $h(t) = -e\hbar^{-1}E_{Dc}t + h_0$. Thus, this evolution is periodic with a Bloch frequency corresponding to the time required for the quasimomentum to scan a full Brillouin zone. It leads to the exchanges

$$\varphi \to h(t)a = -\Omega_B t + \varphi_0 \qquad (24)$$

$$\Lambda(\varphi) \to \Lambda(t) = \frac{\Omega_R}{2(t_a - t_b)\cos(\Omega_B t - \varphi_0) + \sqrt{4(t_a - t_b)^2 \cos^2(\Omega_B t - \varphi_0) + \Omega_R^2}} \qquad (25)$$

$$\Omega(\varphi) \to \Omega(t) = \sqrt{4(t_a - t_b)^2 \cos^2(\Omega_B t - \varphi_0) + \Omega_R^2} \qquad (26)$$

where $h_0 = \varphi_0 a$.

Equations (24)-(26) allow to analyze the spectrum of the tunneling current in the BO regime. Following quasiclassical concept, the tunneling current may be considered as RO with modulated amplitude and frequency. As it leads from equations (24)-(26), the spectral presentation of tunnel current is given by

$$J_T(t) \Rightarrow \sum_{m=-\infty}^{\infty} a_m e^{i(\bar{\Omega} - m\Omega_B)t} + c.c. \qquad (27)$$

where $\bar{\Omega} = (2\pi)^{-1} \int_0^{2\pi} \sqrt{4(t_a - t_b)^2 \cos^2 \varphi + \Omega_R^2} d\varphi$, and $a_m$ are the spectral amplitudes. Below we consider the amplitude relations in the current spectra basing on the numerical solution of motion equations. Here, we note that the small detuning for which $\bar{\Omega} - m\Omega_B \approx \delta\Omega_B$, with $\delta \ll 1$, associated with special type of resonance motion. The above represents a superposition of harmonic oscillations, which one evolves slowly in time with a period $2\pi/(\delta\Omega_B) \gg T_{B,R}$, where $T_{B,R}$ are the Rabi and Bloch periods, respectively. Evaluating the particle position as $x(t) \cong \int_0^t J_T(\tau)d\tau$, we obtain $x(t) \cong (\delta\Omega_B)^{-1} e^{i(\bar{\Omega} - m\Omega_B)t} + c.c.$. It shows the strong enhancement of spatial oscillations, which is similar to the super-Bloch oscillations recently predicted and experimentally observed in optical lattices with cold atoms [40]. In contrast with super-Bloch oscillations, in our case exist two tunneling channels (with respect to the different zones). Thus, the influence of ac field doesn't add up to renormalization of tunneling factors, but leads to their mutual influence via Rabi-oscillations.

## 4. Ultrastrong coupling: results and discussion

In this section we will consider the dynamics and spectra for the case of ultrastrong coupling. As it was mentioned above, the equations of motion are analytically unsolvable and were integrated numerically with the technique [64].

### 4.1. The case of small Bloch-frequency

In this subsection we consider the case of Rabi-frequency strongly exceed Bloch-frequency, while Rabi-frequency is comparable with Bloch-frequency ( $\Omega_R \gg \Omega_B$, while $\Omega_R \cong \omega_0$). The spatial-temporal behavior of the tunnel current for different types of relative values of tunneling factors at ground and excited states is presented in Figures s 3,4. The current oscillates with Bloch frequency. The amplitude for the case $t_a = t_b$ exceed the case $t_a \gg t_b$. The current amplitude is high frequency modulated. In general, the scenario looks like as in the case of the strong coupling.

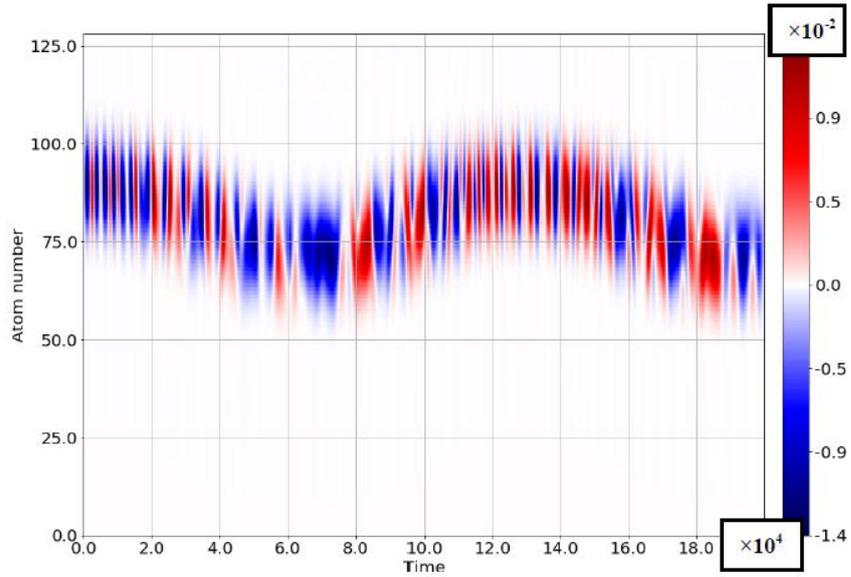

**Figure 3.** Space-time distribution of the tunnel current density with RO and BO in the separated frequency ranges. Here, the quantum transition frequency is taken as the frequency unit, detuning is equal to zero $(\omega = \omega_0)$, $\Omega_B = 0.04$, $\Omega_R = 0.8$, $t_a = 0.4$, $t_b = 0.04$, interatomic distance $a = 20nm$. The initial state of the chain is an excited single Gaussian wave packet $a_j(0) = g\exp\left[-(j-j')^2 a^2/\sigma^2\right]$, $b_j(0) = 0$. Gaussian initial position and width are $j' = 80$, $g = 20$, respectively. Number of atoms $N = 128$.

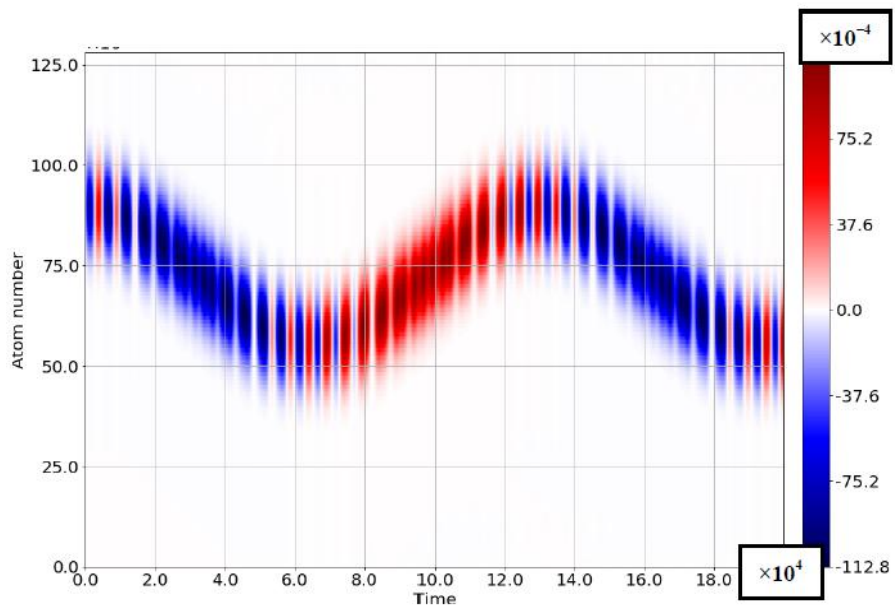

**Figure 4.** Space-time distribution of the tunnel current density with RO and BO in the separated frequency ranges. $t_a = t_b = 0.04$. All other parameters are identical to Figure 3.

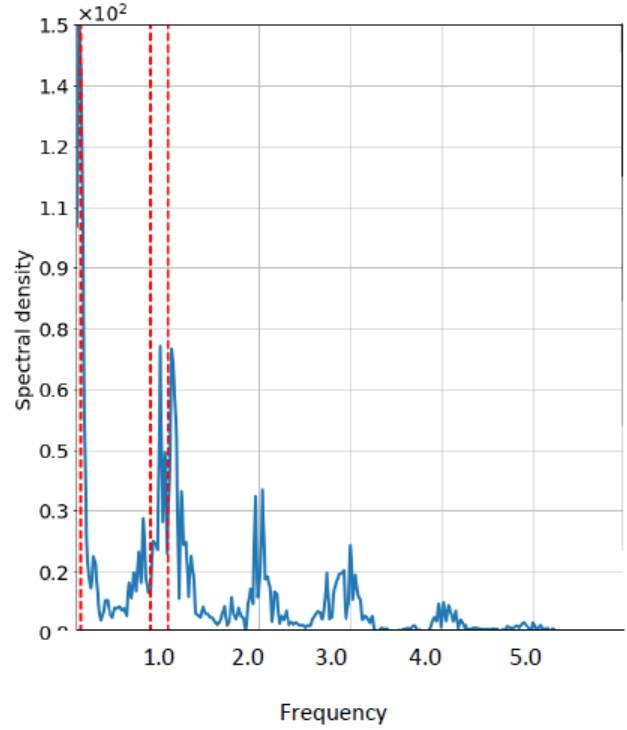

**Figure 5.** Frequency spectrum of the tunneling current density with RO and BO in the separated frequency ranges at the atomic position with number $j = 90$. The quantum transition frequency, Rabi-frequency and Bloch-frequency are denoted by red-colored dashed lines. The quantum transition frequency is taken as the unit. $\Omega_B = 0.04, \Omega_R = 0.8$ All other parameters are identical to Figure 3.

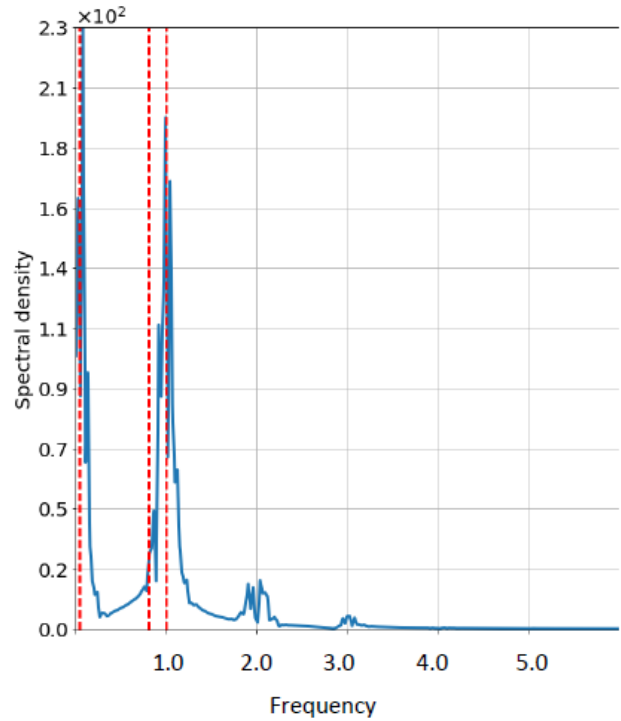

**Figure 6.** Frequency spectrum of the density of the tunneling current with RO and BO in the separated frequency ranges. $t_a = t_b = 0.04$. All other parameters are identical to Figure 5.

In Figures 5,6 there are presented the correspondent spectra. Their qualitative behavior dramatically changes with respect to the strong coupling regime (see equation (27)). The main reason leads from the strong analysis of the system (2),(3), which is free from RWA and based on Floquet theorem. Following it, the probability amplitudes of *j*-th mode are presented in the form

$$\begin{pmatrix} a_{p,j}(t) \\ b_{p,j}(t) \end{pmatrix} = \sum_{m=-\infty}^{\infty} \begin{pmatrix} a_{pm,j}(\varphi) \\ b_{pm,j}(\varphi) \end{pmatrix} e^{-im\omega_0 t} e^{-iv_j(\varphi)t} \tag{28}$$

where $a_{pm,j}(\varphi), b_{pm,j}(\varphi)$ are the periodical functions of the phase shift per unit cell, $v_j(\varphi)$ is the eigen value. Therefore, as it leads from equation (21), the lines $\omega = m\omega_0$ appear at the frequency spectrum and the first of them are really observed at Figures 5,6 in contrast with relation (27). Following the quasi-classical approach, we exchange $\varphi \to -\Omega_B t + \varphi_0$, thus the values $a_{pm,j}(\varphi), b_{pm,j}(\varphi)$ become the time dependent periodic functions with Bloch-period. Such amplitude modulation means appearance the lines $\omega = m\omega_0 + n\Omega_B$, where $m, n = 0, \pm 1, \pm 2, ..., \pm \infty$. The frequency-modulated factors $\exp[iv_j(\varphi)]$ for qualitative analysis may be averaged over the Bloch-period [40]. The dominative support to observable values is defined by two first modes with *j*=1,2. Thus, the discrete part of the spectra in the regime of ultrastrong coupling corresponds to the narrow lines with the frequencies

$$\omega = m\omega_0 + n\Omega_B \pm p\bar{\Omega} \tag{29}$$

where $p = 0, \pm 1$, and

$$\bar{\Omega} = \frac{1}{2\pi} \int_0^{2\pi} \left[ v_1(\varphi) - v_2(\varphi) \right] d\varphi \tag{30}$$

This frequency may be identified with the frequency of RO. The spectrum defined by equation (29) corresponds to the physical picture of RBO as a superposition of three partial motions: i) a single quantum transition from the ground to excited state (or vice versa); ii) the series of periodic quantum transitions between different states with frequency $\bar{\Omega}$ (RO); iii) the spatial oscillations as a comprehensive whole with frequency $\Omega_B$ (BO). Of course, the validity of this picture is strongly limited essentially in view of the fact that the travelling waves used as a model of excitations. But if we go to the wavepacket with strongly defined momentum, this picture will be adequate too, which is supported by the numerical calculations presented below.

We will state only on the analysis and physical interpretation of the dominative lines, which will keep their resolvability in real experimental conditions. The spectra in Figures 5,6 are clear separated to the low-frequency and high-frequency regions. The low-frequency part consist the single line $\omega \approx \Omega_B$, the high-frequency part consist the different sub-harmonics (29). The low-frequency part corresponds to the ordinaryBO; the high frequency spectrum is governed by the RO. The dipole moment and inversion densities are presented at Figures 7,9 and correspondent spectra - at Figures 8,10, respectively.

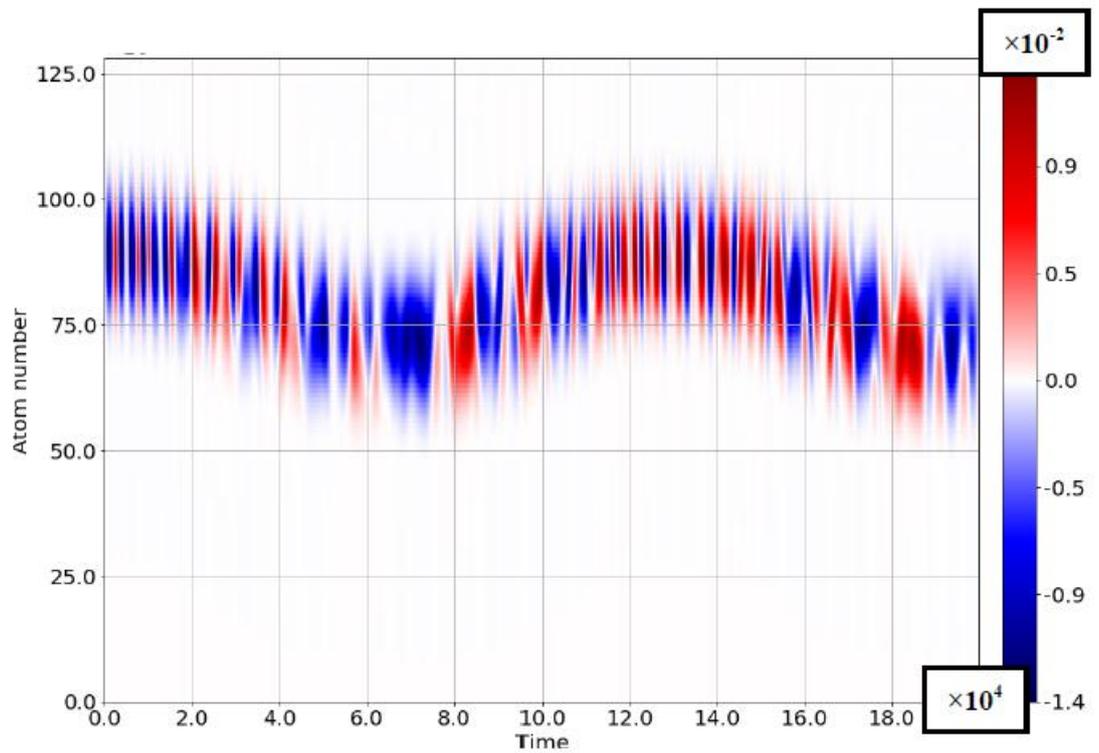

**Figure 7.** Space-time distribution of the dipole moment density with RO and BO in the separated frequency ranges. All parameters are identical to Figure 3.

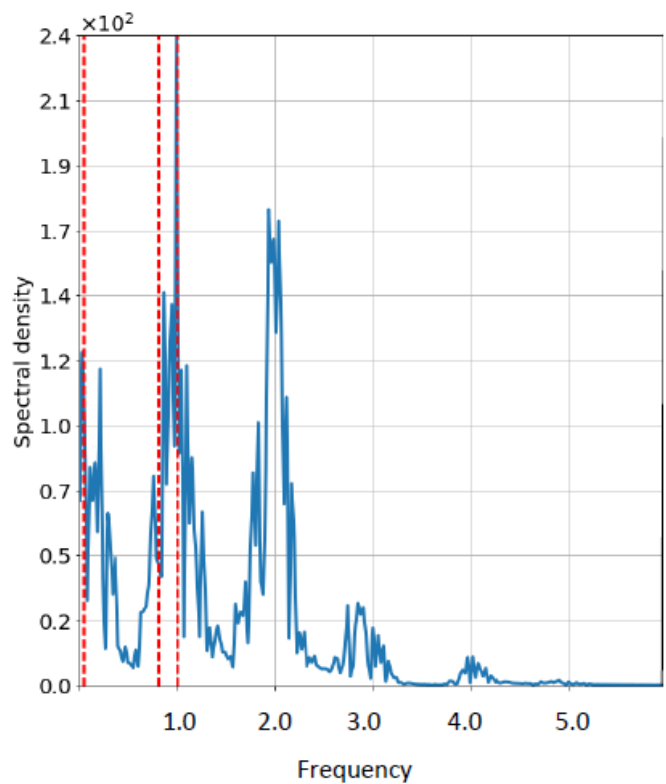

**Figure 8.** Frequency spectrum of the dipole moment with RO and BO in the separated frequency ranges at the atom with position number $j = 90$. All parameters are identical to Figure 3.

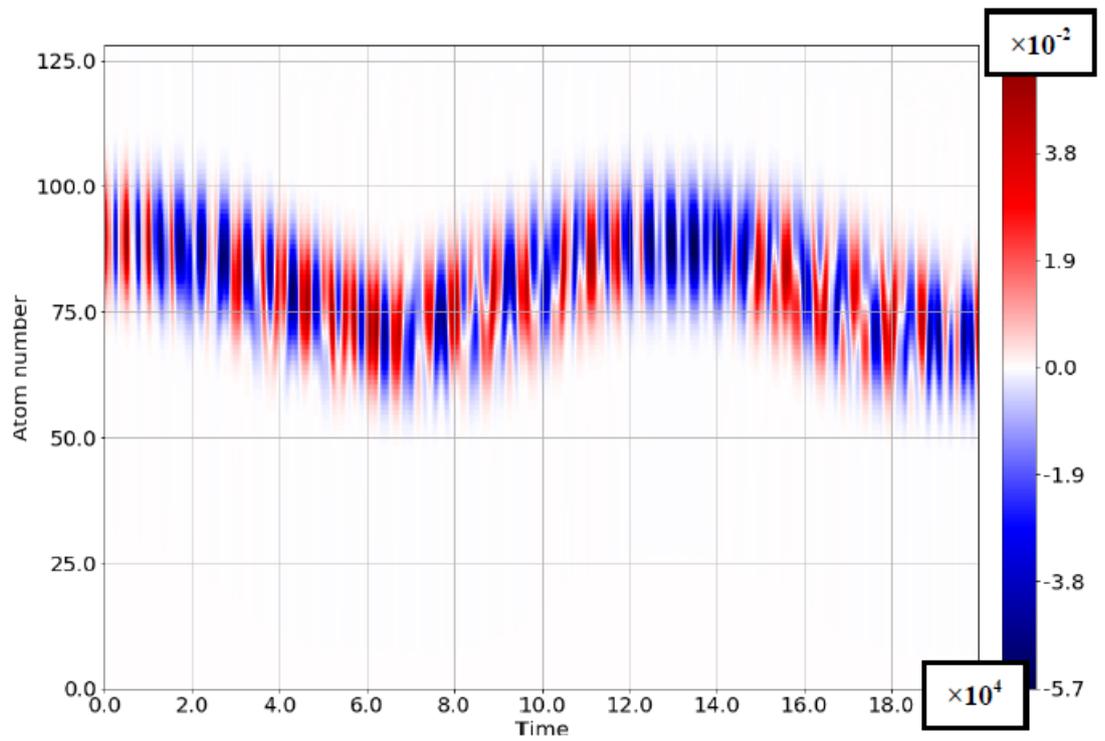

**Figure 9.** Space-time distribution of the inversion density with RO and BO in the separated frequency ranges. All parameters are identical to Figure 3.

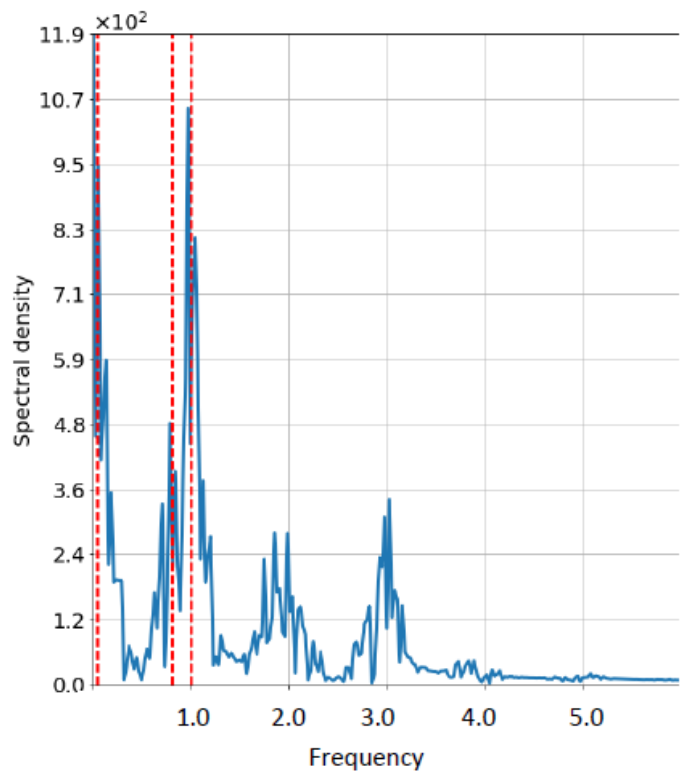

**Figure 10.** Frequency spectrum of inversion density with RO and BO at the separated frequency ranges at the atomic position $j = 90$. All parameters are identical to Figure 3.

As one can see, the dipole current and inversion densities are modulated by the Bloch frequency. The reason is the mutual influence of BO and RO. The low frequency component $\omega \approx \Omega_B$ corresponds to the mutual influence of RO and BO in the so called RBO regime.

*4.2. The case of comparable Rabi- and Bloch frequencies*

In this subsection we consider another case: the Rabi- and Bloch frequencies are comparable. The scenario dramatically changes: the total spectrum is not separable to the low-frequency and high-frequency components. As one can see at Figure. 12, there are appearing the combined sub-harmonics of Bloch and Rabi-frequencies according equation (27) (RBO regime).

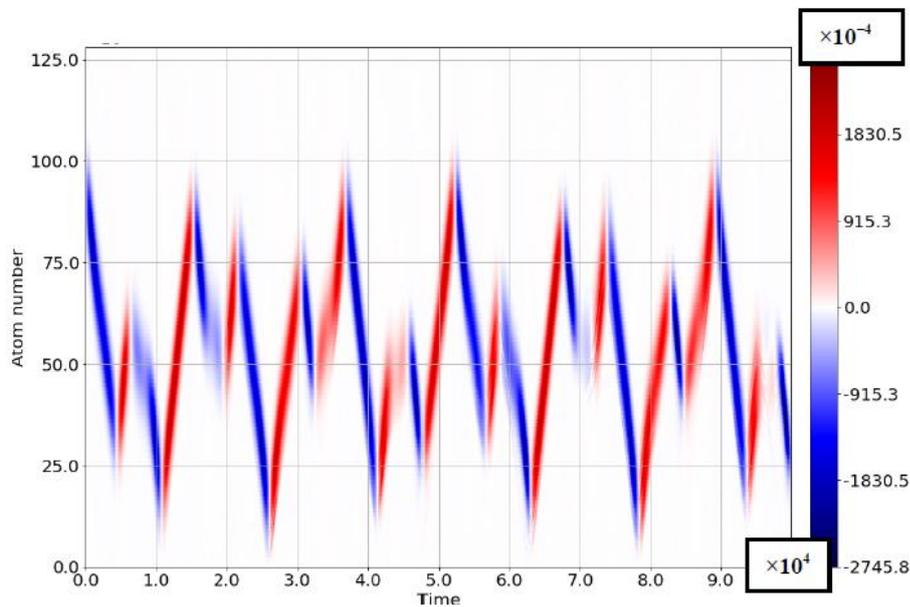

**Figure 11.** The space-time distribution of the tunnel current density with RBO in the ultrastrong coupling regime. Here, the quantum transition frequency is taken as the frequency unit, zero detuning $(\omega = \omega_0)$, $\Omega_B = 0.7$, $\Omega_R = 0.8$, $t_a = t_b = 7.0$, interatomic distance $a = 20 nm$. The initial state of the chain is an excited single Gaussian wave packet $a_j(0) = g \exp\left[-(j-j')^2 a^2/\sigma^2\right]$ $b_j(0) = 0$. Gaussian initial position and width are $j' = 80$, $g = 20$, respectively. Number of atoms $N = 128$.

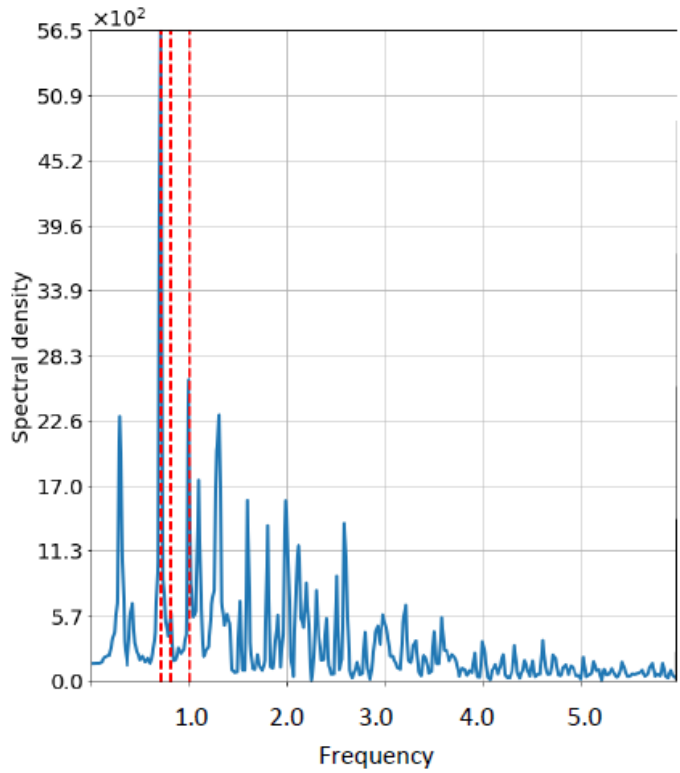

**Figure 12.** The frequency spectrum of tunnel current density with RBO in ultrastrong coupling regime at the atomic position $j = 60$. All parameters are identical to Figure 11.

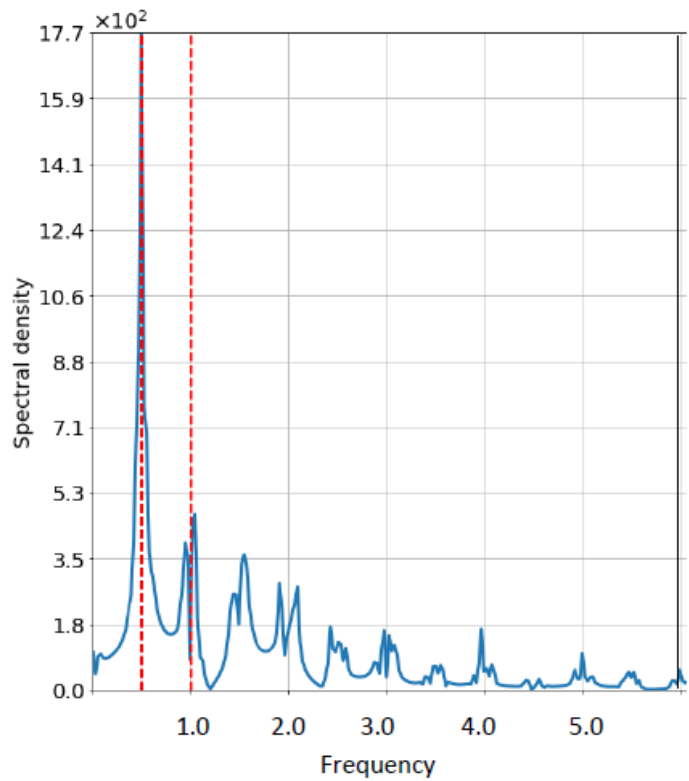

**Figure 13.** The frequency spectrum of the tunnel current density with RBO in the ultrastrong coupling regime at the atomic position $j = 60$. $\Omega_R = \Omega_B = 0.5$, $t_a = t_b = 5.0$. All other parameters are identical to Figure 11.

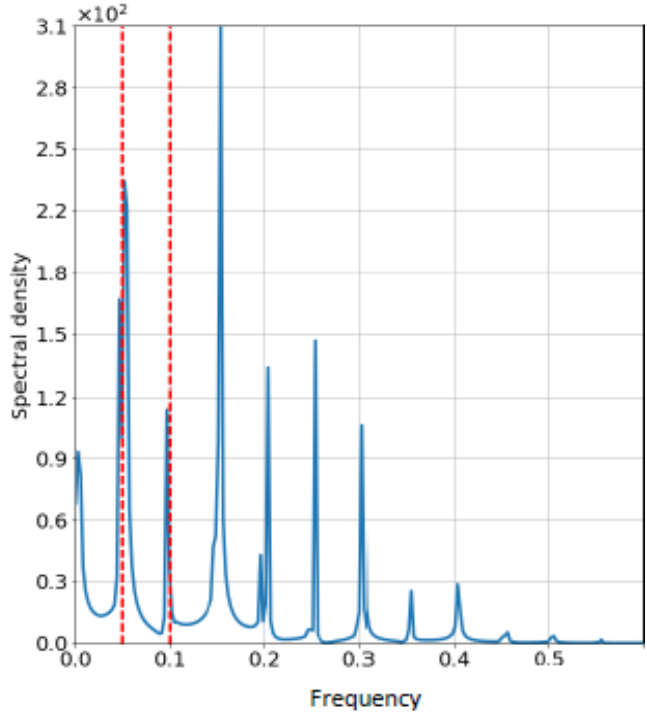

**Figure 14.** The frequency spectrum of the dipole moment density with RBO in the ultrastrong coupling regime at the atomic position $j = 60$. $\Omega_R = \Omega_B = 0.5$, $t_a = t_b = 5.0$.

In the Figures 13, 14 there are presented the spectra of tunneling current and dipole moment densities for $\Omega_B = \Omega_R = 0.5\omega_0$. In this special case a lot of sub-harmonics become degenerate. As a result, the density of spectral lines decreases and grows the ability of their resolution in real experiments avoiding their strong overlap via the dissipative broadening. Appearing the lines $\omega = \omega_0$ $\omega = \omega_0 \pm \Omega_R$, from the first point of view may be associated with Mollow triplet, but no rather convincing arguments for such interpretation exist. The reason is the degeneracy of the modes of different origin in our case, in contrast with Lorentz single-oscillator lines in Mollow triplet [1].

*4.3. The role of the losses*

One of distinctive features of ultrastrong coupling is a comparable value of inter-band transition with radiative and dissipative damping value. Thus, the dissipative and radiative broadening leads to the confluence of neighboring spectral lines. As a result, the discrete lines will interchange with areas of continuous spectra. This discrete lines will have non-Lorentz asymmetric form. It means, that the simple lossless model is not applicable for detail analysis of the spectra thin structure – it allows only to predict the spectral line position and their relative amplitudes. For estimating the role of the losses, we will use the simple phenomenological model used by Scully et all for the analysis of population trapping, lasing without inversion and electromagnetically induced transparency [1]. This model assumes that two levels decay at a rate $\gamma$ [1,65]. It means that the ground state is not really ground: it corresponds to the excited state at the lowest energy level. The wavefunction for such model is given by

$$|\Psi(t)\rangle = \frac{1}{\sqrt{N(1+\Lambda^2)}} \sum_p \left(A_p(t)|a_p\rangle + B_p(t)|b_p\rangle\right) e^{-i\mu t} e^{-\gamma t/2} \qquad (31)$$

where $A_p(t), B_p(t)$ are the same coefficients as in the lossless system.

The tunneling current and its frequency spectrum for Q=15 are shown in Figures 15,16. One can see that the oscillation process quickly reduces (Figure 15). The number of the resolved spectral lines decreases with decay appearance (Figure16) because of the large number of lines run into due to the broadening. The main lines are kept, however their positions are shifted and amplitudes are decreased with respect to the lossless case. Such qualitative behavior agrees with general principles of the oscillations theory [66]. The detailed description of losses must base on the concept of open quantum systems [67] and master equation technique especially adopted to the case of ultrastrong coupling [68]. It goes beyond the subject of this article and may be one of the topics for future research activity.

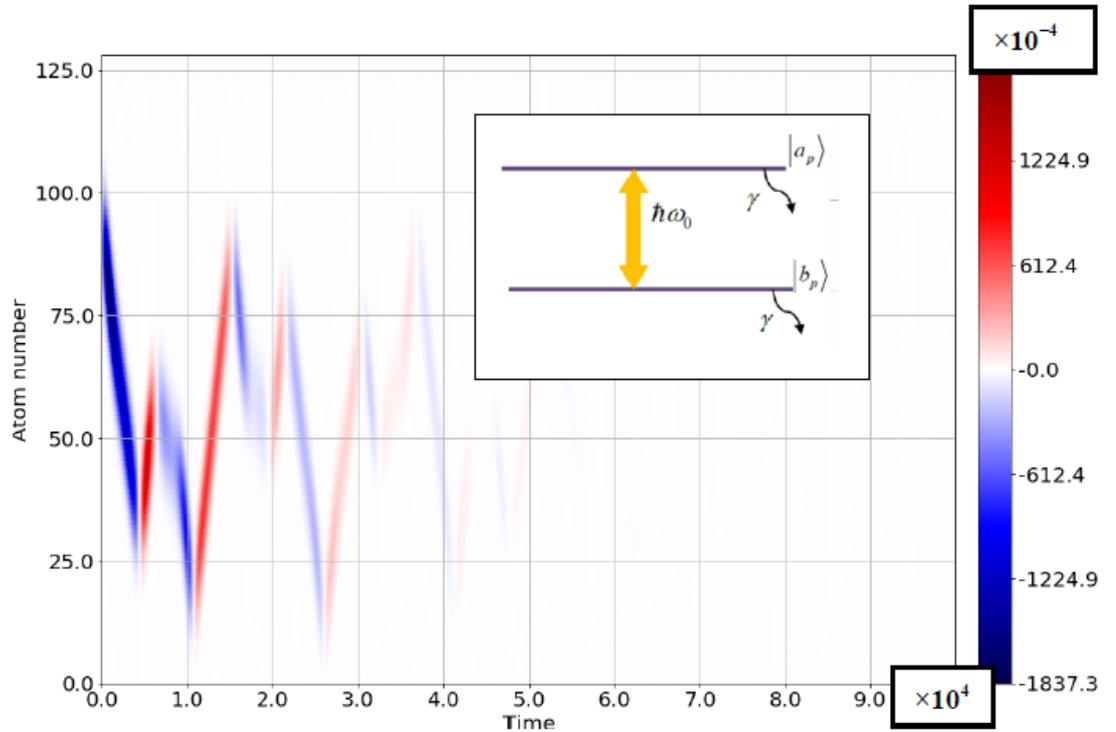

**Figure 15.** The space-time distribution of tunneling current density with RBO in the ultrastrong coupling regime with dissipation. Insert: the energy spectrum of the single atom in the chain. The model assumes that two levels decay at a rate $\gamma$ due to the spontaneous emission. Thus, the ground state is not really ground: it corresponds to the excited state at the lowest energy level. The dissipation property is characterized by Q-factor defined as $Q = \omega_0/2\gamma$. The realistic values for ultrastrong coupling case Q=15-30 [43]. Here, the quantum transition frequency is taken as the frequency unit, detuning is zero $(\omega = \omega_0)$, $\Omega_B = 0.7$, $\Omega_R = 0.8$, $t_a = t_b = 7.0$, interatomic distance $a = 20 nm$, Q=15. The initial state of the chain is an excited single Gaussian wave packet $a_j(0) = g\exp\left[-(j-j')^2 a^2/\sigma^2\right]$ $b_j(0) = 0$. Gaussian initial position and width are $j' = 80$, $g = 20$, respectively. Number of atoms $N = 128$.

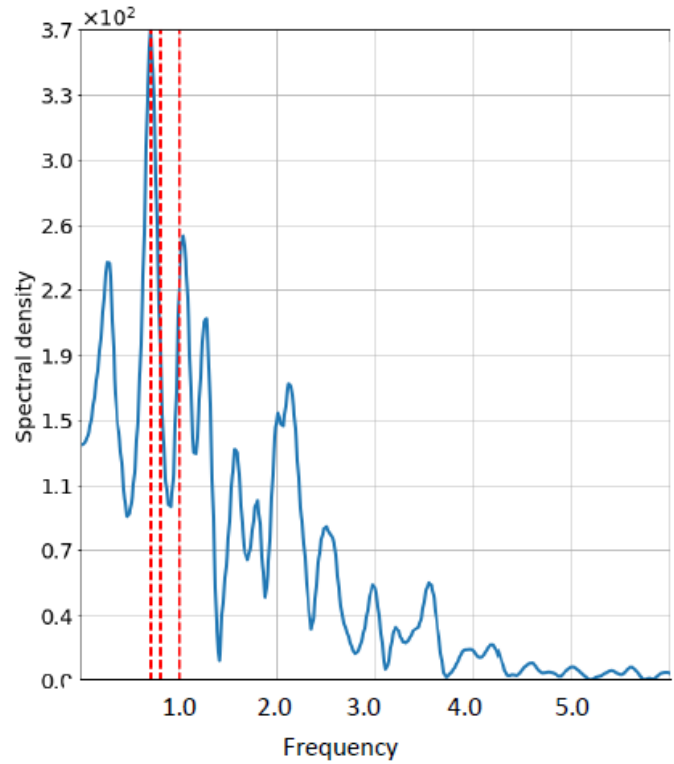

**Figure 16.** The frequency spectrum of tunneling current density with RBO in the ultrastrong coupling regime with dissipation in the atomic position $j = 60$. All other parameters are identical to Figure 15.

*4.4. Stark-effect in one-dimensional atomic chains*

      In this section we consider the 1D chain of two-level atoms driven by dc field only. For a single atom the interaction with dc field leads to the well-known Stark-effect, which consists in the shifting of energy levels [69]. For reachable field values the shift is small and considered as a perturbation of the energy spectrum (weak coupling regime) [69]. The dc field couples all eigen states of the atom, thus the multi-level model with summation over the complete set of atomic terms is applicable to the real atoms. Here, we show that Stark-effect manifests itself in the 1D chains too, but its qualitative picture differed from the single-atomic case. The main reason is the perform ability of ultrastrong coupling, on which we focus below. The system under consideration is shown at Figure17 The dipole moments of the atoms assumed to be oriented orthogonally to the chain axis, the dc field is directed under the angle $\alpha$ with respect to the axis.

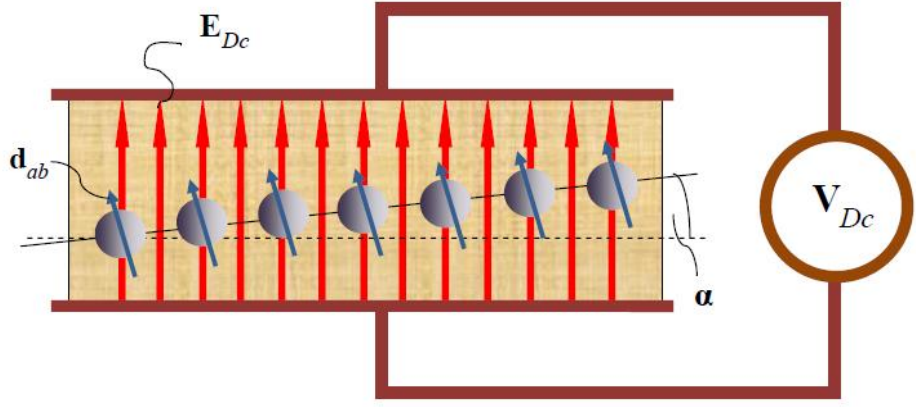

**Figure 17.** General illustration of the periodic two-level atomic chain used as a model, indicating Stark effect in the ultrastrong coupling regime. The chain is placed obliquely between the armatures of plane capacitor with applied dc voltage. The dipole moments (blue arrowheads) oriented orthogonally to the chain axis. Dc field (red arrowheads) oriented angularly with respect to the chain axis. The choice of the angle allows obtaining the arbitrary relative value between Rabi- and Bloch frequencies.

The analytical model in general is similar to the RBO-model developed and used before. The wave-function is given by

$$|\Psi(t)\rangle = \sum_p \left( a_p(t)|a_p\rangle + b_p(t)|b_p\rangle \right) \quad (32)$$

where the probability amplitudes satisfy the system

$$i\frac{\partial a_p}{\partial t} = \left(\frac{1}{2}\omega_0 - p\Omega_B\right)a_p + t_a\left(a_{p+1} + a_{p-1}\right) - \Omega_R b_p \quad (33)$$

$$i\frac{\partial b_p}{\partial t} = -\left(\frac{1}{2}\omega_0 + p\Omega_B\right)b_p + t_b\left(b_{p+1} + b_{p-1}\right) - \Omega_R a_p \quad (34)$$

where $\Omega_R = e\hbar^{-1}d_{ab}E_{Dc}\sin\alpha$, $\Omega_B = e\hbar^{-1}aE_{Dc}\cos\alpha$, are the Rabi- and Bloch frequencies respectively. In contrast with RBO these two frequencies defined by the same field (dc), but their relative value may be arbitrary via the correspondent choice of the field direction (angle $\alpha$).

Let us assume that

$$\Omega_R \approx \omega_0 \quad (35)$$

and

$$\Omega_R \gg \Omega_B \quad (36)$$

Equation (35) corresponds to the resonant condition. It is counterintuitive resonance, which don't coupled with any oscillating forces and represents the attribute of ultrastrong coupling regime. This resonance legitimates the using of two-level model in contrast with the single-atomic case. The inequality (36) means the high amplitude of BO. It allows us to separate the inter-atomic and intra-atomic motions and describe the last one in terms of quasi-classical approximation. The wave-function of inter-atomic motion is given as a superposition of eigen states (12),(13). The account of intra-atomic motion is made via exchange (24). As a result, the total motion is described by the wave-function (15) with

$$\Omega(t) = \sqrt{(t_a - t_b)^2 \cos^2(\Omega_B t - \varphi_0) + \Omega_R^2} \qquad (37)$$

$$\Lambda(t) = \frac{\Omega_R}{(t_a - t_b)\cos(\Omega_B t - \varphi_0) + \sqrt{(t_a - t_b)^2 \cos^2(\Omega_B t - \varphi_0) + \Omega_R^2}} \qquad (38)$$

The application of this solution to the relation of tunneling current (21) leads to the spectrum (29) in the case of Stark effect.

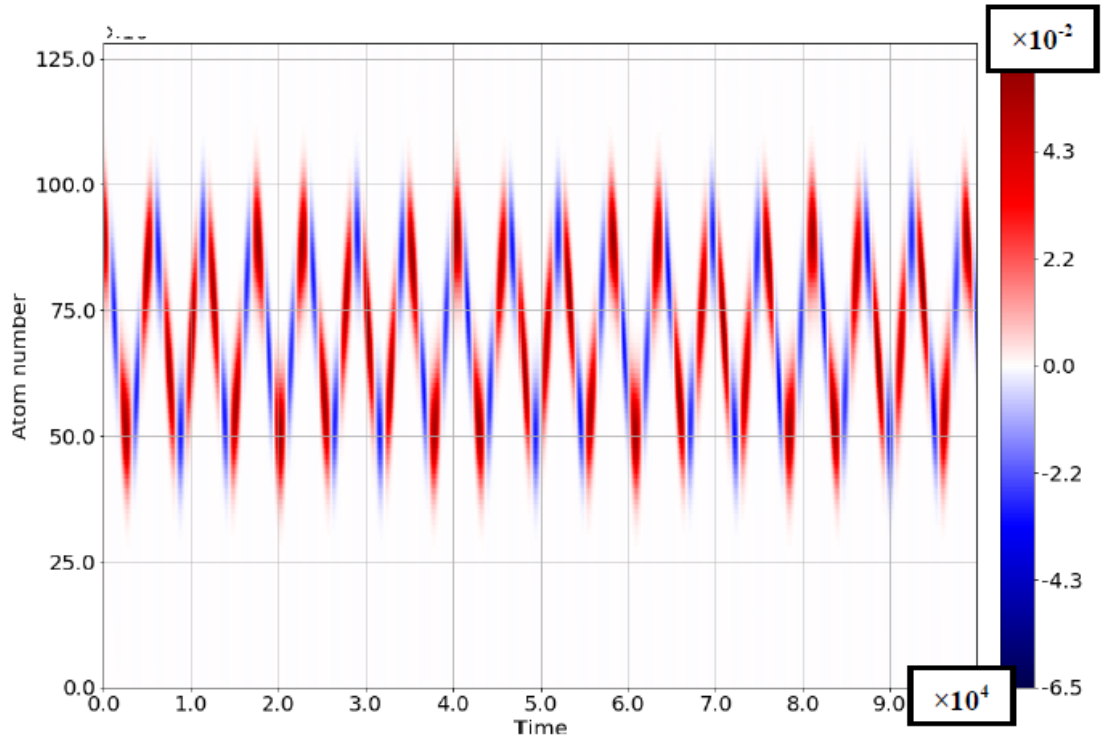

**Figure 18.** Space-time distribution of inversion density with Stark effect. The initial state of the chain is an excited single Gaussian wave packet $a_j(0) = g \exp\left[-(j-j')^2 a^2/\sigma^2\right]$, $b_j(0) = 0$. Gaussian initial position and width are $j' = 80$, $g = 20$, respectively, $\Omega_B = 0.9\omega_0$, $t_a = t_b = 9.0$, number of atoms $N = 128$.

Another approach consists in the numerical solution of equations (33),(34), which is simplified by the absence of oscillation coefficients. The typical behavior of space-time distribution of inversion and tunneling current, as well as the frequency spectrum of tunneling current are shown in Figures 18-20. The scenario of the process may be explained by the next way. The ordinary Stark effect for the single atom consists in the perturbed energy spectrum and mixing of the states due to the dc field [69]. In our case the situation is similar, however the energy shifting is not small. The system is initially prepared in the one of non-perturbed states, which is not relates to the eigen-modes of the system with the dc field applied. As a result, the new eigen-modes become coupled, and we observe their mutual oscillations in inversion with Rabi-frequency (see Figure 18). These oscillations propagate along the chain with Rabi-period due to the tunneling and produce the intra-zone current (Figure 19). The value of this current strongly increases due to the resonant condition (35). The spectrum of this currents is a set of lines $\omega = n\Omega_R$, $n=1,2,...$ (see Figure 20). In the opposite case $\Omega_R \ll \Omega_B$ (dc field directed predominately along the chain) the inter-zone coupling becomes small, which leads to the suppression of intra-atomic motion. Thus, the total motion adds up to the oscillative inter-atomic tunneling, which corresponds to the ordinary BO.

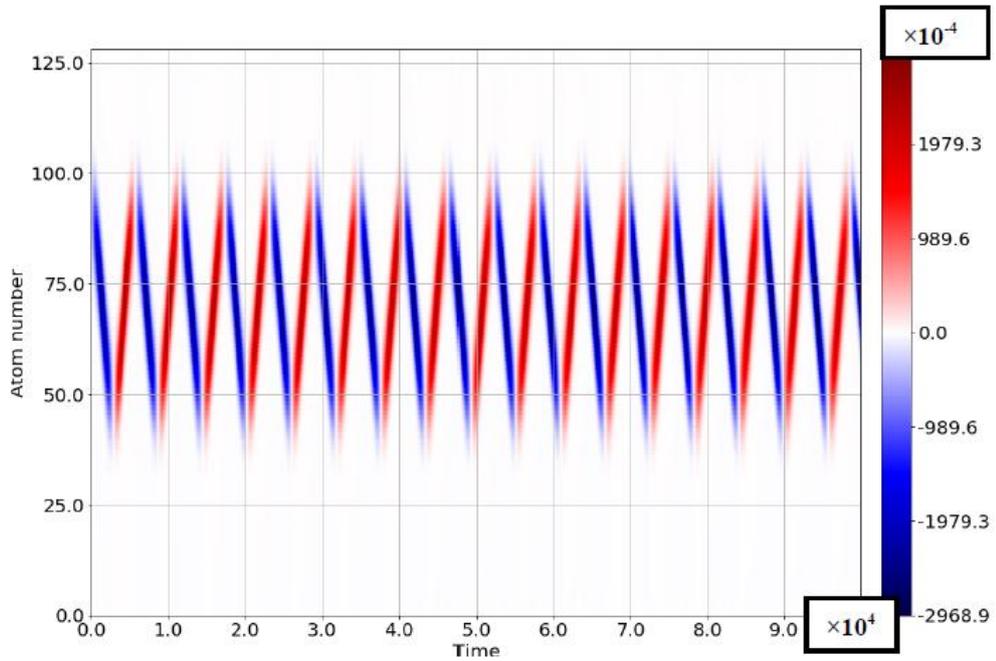

**Figure 19.** Space-time distribution of tunneling current density with Stark effect. All parameters are identical to Figure 18.

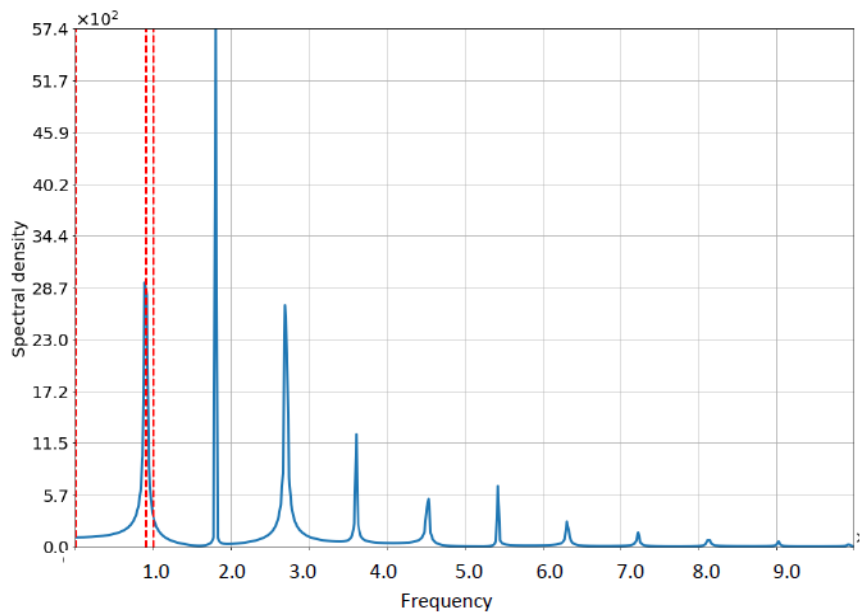

**Figure 20.** The frequency spectrum of tunneling current density with Stark effect. All parameters are identical to Figure 15.

## 5. Conclusion

In summary, we developed the physical model of 1D two-level atomic chain with the tunneling inter-atomic coupling simultaneously driven by dc and ac fields in the regimes of strong and ultrastrong coupling. The approximate analytical solutions and results of computer modeling of Rabi-Bloch oscillations have been presented and discussed. There are considered also such limited cases as Stark-effect, Rabi-waves, Bloch oscillations. The temporal dynamics as well as the frequency spectra of tunneling current have been presented. It was shown that the passage to the ultrastrong coupling dramatically changes the scenario of Rabi-Bloch oscillations. It manifests in appearance of additional lines in the spectra of the tunneling current. These lines are identified as a contribution of anti-resonant interactions and remain even in the case of high losses character for ultrastrong coupling. The promising potential applications of obtain results in novel type of THz spectroscopy for nano-electronics have been proposed.

**Author Contributions**

Developments of the physical models, derivation of the basis equations, interpretation of the physical results and righting the paper have been done jointly. The MATHEMATICA Numerical Python calculations and figures were produced by I. L. with [70,71].

**Conflicts of Interest**

The authors declare that they have no conflict of interest.